# Shape transition in ZnO nanostructures and its effect on blue-green photoluminescence

Manoranjan Ghosh[1] and A.K. Raychaudhuri[2]
*DST Unit for Nanoscience, S.N.Bose National Centre for Basic Sciences, Block-JD, Sector-3, Salt Lake, Kolkata-700 098, INDIA*

[1] Email: mghosh@bose.res.in
[2] Email: arup@bose.res.in

**Abstract:** We report that ZnO nanostructures synthesized by chemical route undergo a shape transition at ~ 20 nm from spherical to hexagonal morphology thereby changing the spectral components of the blue- green emission. Spherically shaped nanocrystals (size range 11 – 18 nm) show emission in the range of 555-564 nm and the emission shifts to the longer wavelength as the size increases. On the other hand, rods and hexagonal platelets (size range 20-85 nm), which is the equilibrium morphology after the shape transition, show emission near 465-500 nm and it shifts to shorter wavelength as the size increases. The shape transition also leads to relaxation of microstrain in the system. Our analysis shows that the visible emission originates from a defect layer on the nanostructure surface which is affected by the shape transition. The change in the spectral component of the blue green emission on change of shape has been explained as arising from band bending due to depletion layer in smaller spherical particles which is absent in the larger particles with flat faces.

## 1. Introduction

The room temperature photoluminescence (PL) of ZnO nanostructures typically exhibits a sharp emission in the near UV range (originating from excitonic mechanism) and a broad emission band in the visible region of the spectrum which is linked to defects in ZnO [1]. Different types of defects have been proposed to explain the two commonly observed defect emission bands in the visible range – a blue-green band ($\lambda_{bg}$ ~ 500-560 nm) and an orange-yellow band ($\lambda_{oy}$ ~ 610-650 nm) [2, 3]. The specific identification of the nature of defects for the emission in the visible region is difficult because of simultaneous existence of various types of defects in ZnO nanostructures which often depend on the method of preparation [4]. Fundamental questions like, evolution of the visible emission with the nanostructure size and morphology remain unresolved [5]. In an effort to correlate size, defects and PL of ZnO nanoparticles [6], it has been found that blue shift of the green emission for increasing size (from 25 -75 nm) is opposite to that observed on nanowires of larger diameters ranging from 50-300 nm [7]. In a recent report, the blue green emission of ZnO nanostructures grown by Pulsed Laser Deposition was found to exhibit blue shift and reduced intensity when the size increases and the morphology was found to change from spherical to rods and hexagonal plates [8]. However, the origin of this anomalous shift in emission energies has not been explained and the role of morphology of the nanostructures on the visible emission has not been investigated.

Our investigation carried out on size and morphology controlled ZnO nanostructures prepared by chemical method shows that a clear shape transition occurs as the size increases beyond ~20 nm. The shape change leads to a sharp change in the spectral content of the blue-green visible emission. Nanostructures with spherical shape (10 – 20 nm) show emission near 550 nm whereas rods and hexagonal platelets with flat faces emit near 500 nm. Thus the problem of the anomalous shift in visible emission on size variation, stated in the paragraph above, can be related to the issue of change in shape on size reduction. Both these emissions (around 500nm and 550nm bands) show very systematic dependence on the surface to volume ratio which confirms the surface related origin of these emissions. We show below that the emission from ZnO nanostructures is strongly dependent on the band bending effects arising due to the presence of positive surface charge on the surface of ZnO nanostructures. The shape transition leads to a change in the band bending effect which in turn changes the emission spectrum.

## 2. Experimental details

The controlled ZnO nanostructures (having wurtzite crystal structure) have been synthesized by acetate route described in detail elsewhere [9, 10]. The samples of different sizes were prepared in an autoclave by the same chemical route under 54 atm. pressure at 230 $^0$C by varying the growth time. By this method one can obtain particles of different morphology with a wide range of sizes. Precursor concentrations and reaction time have been increased to synthesize larger particles. The details of representative nanostructures are given in Table 1. To synthesize the samples of different sizes, varying amount of $Zn(CH_3COO)_2 \cdot 2H_2O$ has been reacted with NaOH ( concentration three times that of the zinc acetate concentration) for duration ranging from 0.5 hour to 6 hours. Two groups of samples have been investigated. Gr-I samples in Table 1 with spherical morphology of sizes 11 nm, 14 nm and 18 nm are grown in the autoclave for a duration of 0.5 h, 1 h and 1.5 h respectively. Gr-II samples, having hexagonal and rod like morphology with size>20 nm are prepared by allowing longer growth time (up to 6 h) in the autoclave. Thus the sample preparation has been done using the same method and chemistry while the size variation has been done by changing only the time of growth within the autoclave.

The nanostructures were imaged and their lattice images were taken with a 200 kV HRTEM Jeol Electron Microscope (Model No. JEM2011). The X-ray diffraction (XRD) data were collected by a Philips X'pert Pro X-ray diffractometer operated at a voltage 45 kV and current 40 mA. Room temperature optical properties were studied by photoluminescence spectrometer (Jobin Yvon, FluoroMax-3) where a xenon arc lamp has been used as illumination source. The Zeta potentials were measured at 20$^0$C using Malvern Instruments' Zetasizer systems Nano ZS.

# 3. Results

## 3.1 TEM and XRD study

Representative transmission electron microscope (TEM) images of the samples are shown in figure 1. We find that the samples with size below ~20 nm have a spherical shape. However, for sizes > 20nm there is a shape transition and these samples show mostly hexagonal facets and some of them show rod like morphologies. Representative TEM images of spherical ZnO nanostructures of average sizes 10 nm and 16 nm are shown in figure 1 (a) and (b), grown for 0.5 h and 1.5 h duration within the autoclave respectively. Magnified view of the samples in (a) is given in its inset while that of the samples in (b) are shown in the figure (c). Samples with rod like and hexagonal morphologies were obtained by growing for a longer time and have been shown in figures 1 (d) and (e) respectively. Lattice image [figure 1(f)], indexed SAED pattern [figure 1(g)] and FFT of the lattice image [figure 1(h)] of nanostructures shown in (e), confirms their single crystalline nature as well as their atomically flat face. The magnified view of a single rod is shown in the inset of figure 1 (e). The TEM data (Figure 1 and Table 1) clearly establishes that spherical ZnO nanoparticles prepared by chemical routes undergo a shape change leading to hexagonal morphology with sharp edges (often rod like) when the size exceeds the value of 20 nm.

The crystal structures and sizes of all the samples have been analyzed by X-ray diffraction technique. Representative X-ray diffraction data for the samples investigated are indexed in figure 2 which confirms wurtzite symmetry of the synthesized nanostructures. The size determined by TEM and XRD analysis (listed in table 1) agrees well within the range of ± 2 nm for the nanostructures of size below 25 nm. The root mean square (rms) distribution in size ($d$) of the samples can be quantified as $\Delta d_{rms}/d \approx$ 10% for samples (i) and (iv) in table 1 and $\Delta d_{rms}/d \approx$ 20% for rest of the samples investigated in this work.

**Table 1.** Size and morphology of ZnO nanostructures

| Identification | | Morphology | Size by TEM (nm) | Size by XRD (nm) | $E_{NBE}$ (eV) |
|---|---|---|---|---|---|
| I | (i) | Spherical | 10 | 11 | 3.34 |
|   | (ii) | Spherical | 12 | 14 | 3.35 |
|   | (iii) | Spherical | 16 | 18 | 3.33 |
| II | (iv) | Rod diameter | 24 | 23 | 3.33 |
|   | (v) | Hexagonal platelet | 55 | 44 | 3.29 |
|   | (vi) | Rods and square pillars | 100 | 85 | 3.27 |

The change in shape as observed from the TEM images is accompanied by strain relaxation in the spherical particles. We confirm this by measuring the microstrain from the XRD data. Both microstrain as well as size contributes to the line width in the XRD peak. One can separate out the contributions of both by the Williamson-Hall analysis using simplified integral breadth method [12]. The Integral Breadth (β), defined as the ratio between peak area and the intensity maxima for both size and strain broadened profile is given by [12],

$$\beta^* = \frac{1}{D} + 2.\varepsilon.s \qquad \text{(in sin}\theta\text{ scale)} \qquad (3)$$

where D is the average particle diameter, ε is the strain, $\beta^* = \beta\cos\theta/\lambda$, and $s = 2\sin\theta/\lambda$. Each peak of the XRD data (figure 2) was fitted by Pseudo-voigt function to find out the value of integral breadth (β). The average particle diameter was obtained form the y-intercept of the $\beta^*$ vs. s plot and this matches with that found from the TEM (Table 1) (average diameters determined by XRD and TEM agree within the range of ± 10–20% for all the samples studied). W-H analysis also allows us to find the average strain from the slope of the linear fit of the data points. The strain values calculated by this method were plotted with the size of the nanostructures in figure 3. We note that the microstrain is higher for spherical particles (Gr-I) and then collapses when shape changes for Gr-II samples. Therefore it appears that the shape transition is driven by the relaxation of built in strain in the spherical particles.

The change, from spherical to a shape with flat surfaces occurs due to widely different surface energies of the ZnO crystal faces [13]. The spherical shape will contain some high energy surfaces and in fact can expose in some places the high energy basal planes. This excess energy will be lowered by a transition to a hexagonal (or rod like) shape that will expose the lowest energy {0001} surfaces. Such a surface energy driven shape transition by which the excess energy and the microstrain of the nanostructure is relaxed, is well studied in nanoparticles of many systems [14].

## 3.2 Photoluminescence study and its dependence on shape

The change in shape from spherical to hexagonal (or rod shaped) morphology, as evident from the TEM micrographs, leads to a non-monotonous change in the visible emission energy of their PL spectra (excitation at 325 nm). In figure 4, we show the emission spectra over the entire range. The emission in the range 370-380 nm is the near band edge emission. It can be seen that the blue–green emissions in the range 470 nm-565 nm show a distinct shift in the emission energies as the particle size as well as the shape changes. The visible emission from Gr-I nanoparticles with spherical shape is peaked around 550nm while for Gr-II nanostructures with hexagonal and rod shapes, the emission is centered at 500nm. In this paper we propose that the observed change in the emission energy for ZnO nanostructures as their size increases beyond 20nm is actually related to its change in shape. It is noted that the shape transition (which occurs with size) as well as its impact on their photoluminescence has not been observed earlier in ZnO nanostructures.

Analysis of the broad emission in the blue-green region shows that it is composed of two emission bands [shown as dotted line in figure 4 for sample (iii)] which are marked as P1 and P2 [11]. The spectral content of the blue-green band is determined by the relative weight of the two lines. We find that the change in shape brings about a sharp change in the relative contribution of the two lines. P2 line is dominant in Gr-I samples (size ranging from 10-20nm) having spherical morphology. As faceting starts appearing for nanostructures greater than 20 nm (Gr-II samples), intensity of P2 line gradually reduces and P1 emerges as the dominant emission band. This can be seen in figure 5 (upper panel) where we show the variation of the intensity ratio of the two lines ($I_{P2}/I_{P1}$). The shape change suppresses the intensity of P2 emission and changes sharply the energies of emission of both the bands. Emission energies of P1 ($E_{P1}$) and P2 ($E_{P2}$) also show a sudden rise when the shape transition occurs (middle and lower panel of figure 5). Both $E_{P1}$ and $E_{P2}$ shift to lower energies as particle size increases for Gr-I samples (spherical morphology). In case of Gr-II (rods and hexagonal platelets) samples, the energies of emission of both the lines remain more or less size independent after the sharp change that occurs at the shape transition. The above discussion leads us to the conclusion that the anomalous size dependence in the visible emission of the ZnO nanostructures appears to be related to their morphology.

It is important to note that the emission in the UV range (from 370-380 nm) is an intrinsic property of the wurtzite ZnO and originates due to excitonic recombination. Due to quantum confinement it shifts to higher energy as we reduce the size of the nanostructures. Emission energy of this band edge emission (NBE) obeys an inverse dependence on the diameter (1/diameter), irrespective of the morphology and the method of preparation [Table 1]. Thus the visible emission is specific to the mechanism and the change in visible emission energy indicates the presence of two different mechanisms for Gr-I and Gr-II samples.

To establish that the defect centres responsible for the visible PL are located near the surface, we analyzed the ratio of the intensities of NBE to visible emission ($I_{NBE}/I_{VIS}$) as function of size (figure 6). Increase in this ratio as the size increases shows the decreasing contribution of the visible emission which is expected if the centres responsible for the visible emission are primarily located at the surface. A simple model [7] can be used to explain the intensity dependence on the size and thus obtain the thickness of the surface layer emitting visible light. The model essentially measures relative strength of surface to volume. The intensity ratio of the NBE to visible emission for cylindrical wires (Gr-II) of radius r and with a surface recombination layer of thickness t, is given by [7]

$$\frac{I_{NBE}}{I_{VISIBLE}} = C\left(\frac{r^2}{2rt-t^2}-1\right) \qquad (1)$$

By similar approach we can find out the same ratio for spherical particles (Gr-I) of radius r and a surface recombination layer of thickness t

$$\frac{I_{NBE}}{I_{VISIBLE}} = C\left(\frac{r^3}{3rt(r-t)+t^3}-1\right) \qquad (2)$$

The constant C, along with the other quantities contains the oscillator strengths which in turn depend on the particle morphology, is expected to be different for the two groups of particles. Equations (1) and (2) have been employed to fit the experimental data for Gr-II and Gr-I samples respectively as shown in figure 6. The thickness of the surface recombination layers as obtained from the model is t ≈ 4.0 nm and t ≈ 3.6 nm for Gr-I (spherical) and Gr-II (rods and hexagonal platelets) particles respectively. It is important to note that the thickness of the surface layer (t) is comparable but somewhat higher in case of smaller spherical particles than that in the rods and hexagonal platelets of relatively bigger size. Thus the relative volume associated with surface layer is much higher for small spherical particles compared to the rods and hexagonal nanostructures. The simple model above establishes that the visible emission centres are indeed located in a small region around the surface. This is an important issue because as we will see below the physical location near the surface makes the emission sensitive to surface charges and the band bending that occurs near the surface.

## 4. Discussion

The main result of this paper is the shape transition in the ZnO nanostructure when the size becomes more than ~20nm which severely affects the visible emission arising from the defects located near the surface. In the following we discuss the likely origin of this phenomenon with a model and validate this model with additional experiments. First we discuss the spectral shift of blue-green band on shape change. As shown earlier, the emission band in this region is composed of two lines (P1 and P2). It has been shown through figures 4 and 5 that the change in the spectral content occurs because of suppression of the P2 line relative to the P1 line when the nanostructures with flat faces emerge. It has also been shown before that there is a surface region from where the defect related emission occurs.

Based on the suggestion of reference [11], in figure 7 we show the proposed origin of the two emission lines P1 and P2. In ZnO the defect responsible for visible emission in this wave-length range is oxygen vacancy denoted as $V_O$ [16]. The two lines originate depending on whether the emission is from doubly charged vacancy centre $V_O^{++}$ (P2) or singly charged vacancy centre $V_O^{+}$ (P1). The suggested emission process is shown schematically in figure 7. The $V_O^{++}$ centre, created by capture of a hole by the $V_O^{+}$ centre in a depletion region, leads to emission in the vicinity of 2.2 eV ($E_{P2}$), which is the P2 line. The singly charged centre ($V_O^{+}$) in absence of a depletion region becomes neutral centre ($V_O^{X}$) by capture of an electron (n-type ZnO) from the conduction band which then recombines with a hole in the valence band giving rise to an emission (P1) at 2.5 eV ($E_{P1}$). The shift in the visible emission is thus related to the change in the relative strengths of the two emission bands. Now we show that the relative occupation of the two charged centres ($V_O^{++}$ and $V_O^{+}$) is related to the band bending at the depletion layer near the surface which in turn is related to the surface charge of the nanostructures.

In spherical nanoparticles of small size, there will be appreciable depletion layer near the surface [11]. The thickness of the depletion layer depends on the density of native carrier concentration in the nanoparticles. The depletion layer will make the surface of the ZnO nanoparticles positively charged (inherently ZnO is n-type). In ZnO with native defects (n-type), the depletion region has a typical thickness in the range 5-10 nm. Such a depletion layer will give rise to band bending in spherical nanoparticles whose size is comparable to depletion width [15]. The band bending changes the chemical potential in such a way that it populates doubly charged $V_o^{++}$ preferentially by which the P2 emission predominates. A proof for the presence of these $V_0^{++}$ centres is a rather substantial positive zeta-potential (18 mV) measured in these nanoparticles of small size dispersed in ethanol. The shape change leads to hexagonal particles or rods with flat surfaces. These particles/rods have severely reduced zeta-potential (~ 8mV). This reduces the depletion layer and the associated band bending. As a consequence this will change the occupancy of the electron energy levels in these larger faceted particles which decreases the occupancy of the $V_o^{++}$ centers leading to reduction in emission of the $P_2$ band. The occupancy will be like the bulk energy level and the predominant emission will be from the singly charged vacancy ($V_o^+$), namely the P1 line. Thus the shape change suppresses the band bending leading to stronger emission at P1. Therefore the role of shape change is to modify the band bending and the relative occupancy of the $V_o^{++}$ and $V_o^+$ levels which in turn changes the dominant emission channel. The substantial reduction in zeta potential value on shape change is a proof that the surface charge (and the depletion layer) is responsible for this process.

Briefly, the above model is based on the fact that the visible luminescence is linked to the oxygen vacancy and the depletion layer near the surface plays an important role in determining the band bending and the associated consequences in the emission. The shape transition, leads to a change in the surface charge which in turn changes the extent of band bending. The depletion layer which leads to the band bending makes the surface of ZnO nanostructures positively charged as established by the direct measurement of the zeta-potential. So, there is a link between the visible emission and the charge at the nanostructure surface. To establish this relation, we actually modify the surface charge of the nanoparticles by embedding them in electrolytes which leads to a change in their visible emission.

The surface charge driven shift in the relative contribution of the spectral components in the broad visible emission can be seen even in case of nanostructures having the same shape and size by dispersing them in electrolytic solution of varying strength. The positively charged ZnO nanoparticles dispersed in a solution are surrounded by oppositely charged ions. The net (+)ve charge developed at the particle surface affects the distribution of ions in the surrounding interfacial region, resulting in an increased concentration of counter ions (ions of opposite charge to that of the particle) close to the surface. Thus an electrical double layer exists around each particle. So, addition of $LiClO_4$ (a salt of strong base and weak acid) to the dispersion of ZnO nanoparticles in ethanol reduces the Zeta potential at the interface between ethanol and ZnO nanostructures. As we keep on increasing the concentration of $LiClO_4$ in ethanol, the Zeta potential around the surface decreases and finally assumes a negative value (nearly equal to -7 mV for 1 M $LiClO_4$ in ethanol). In figure 8 (a), we plot representative PL spectra of spherical ZnO nanoparticles dispersed in an electrolytic solution having different concentrations of $LiClO_4$ as indicated in the graph. We observe gradual reduction in the overall intensity of the visible emission on increasing the $LiClO_4$ concentration. The broad visible emission has been decomposed into two peaks as in the case of PL spectra of ZnO nanostructure having various shapes and sizes. In figure 8 (a), we show the two decomposed peaks for the particles dispersed in pure ethanol ($P1_{0M}$ and $P2_{0M}$) and 1M $LiClO_4$ solution in ethanol ($P1_{1M}$ and $P2_{1M}$). In figure 8 (b), we plot the intensity ratio between the two spectral components and observe that the relative contribution of P2 band decreases on increasing the $LiClO_4$ concentration. This observation totally agrees with the trend observed in the relative contribution of the two spectral components when the spherical particles grow in size and finally change their shape. Also the spectral position of the two components show a gradual decrease in energy on increasing the electrolyte concentration as observed in the case of spherical particles (Gr-I). For maximum concentration of $LiClO_4$, the observed shift in the spectral position of P2 ($E_{P2}$) is more (0.054 eV) in comparison to that of P1 (0.032 eV). This shows that the P2 emission is more sensitive to the environment. Finally, we observe a similar spectral shift in P1 and P2 emission energy for the spherical particles (Gr-I). Therefore the basic assumption, that the surface charge and band bending is responsible for change in the visible emission in ZnO is validated.

## 5. Conclusion

To summarize, we find that ZnO nanoparticles undergo a shape change at a size ~20 nm from spherical to hexagonal shape. The change in shape is associated with a collapse of the microstrain. It also leads to a sharp change in the blue-green emission band from ZnO nanostructures. The change in spectral nature of the emission band on shape change has been explained as due to change in band bending arising from depletion layer on the surface of nanoparticles. We establish that the depletion layer is linked to the net positive charge residing at the surface on the nanoparticles.

## Acknowledgements
The authors acknowledge the technical help from TEM facility at the Indian Association for the Cultivation of Science, the Department of Science, UGC-CSR Kolkata centre for Zeta potential measurements and Department of Science and Technology, Government of India, for providing the financial support as a unit for Nanoscience.

# Figures

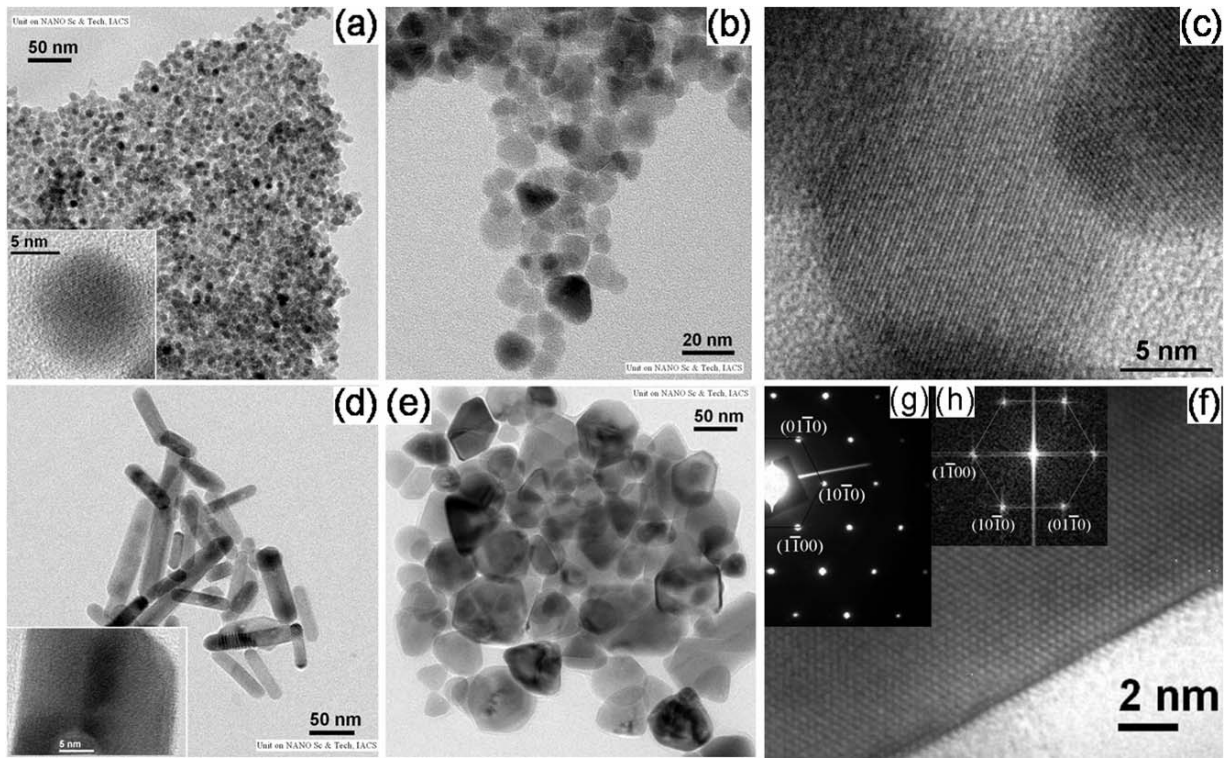

**Figure 1.** *Upper panel:* **TEM images of spherical ZnO nanostructures of average sizes (a) 10 nm and (b) 16 nm. Inset of figure (a) shows the close view of a single particle shown in (a). The high resolution image of nanoparticle in (b) is shown in figure (c).** *Lower panel:* **TEM micrograph of (d) rod like and (e) hexagonal shaped ZnO nanostructures. Inset of figure (d) shows a close view of a single rod. The high resolution lattice image (f), SAED pattern (g) and FFT of the lattice image (h) are presented for the hexagonal particles shown in (e). Sharp boundary of the hexagonal particle can be seen.**

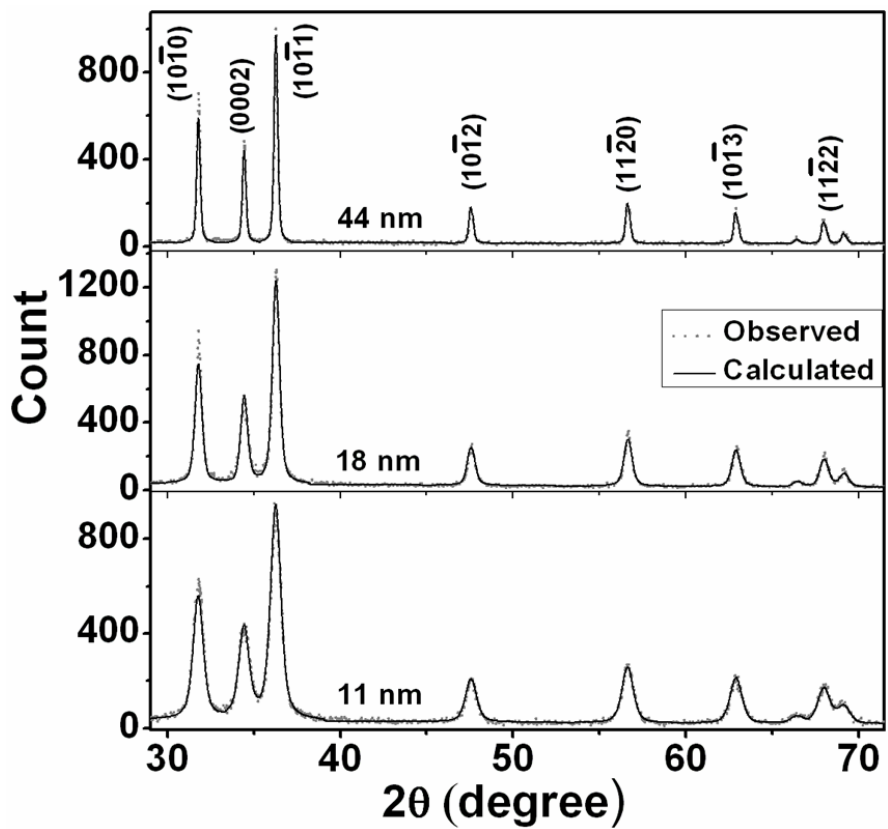

**Figure 2.** XRD data for select samples as indicated on the graph show wurtzite type structure adopting symmetry P63mc (peaks are indexed).

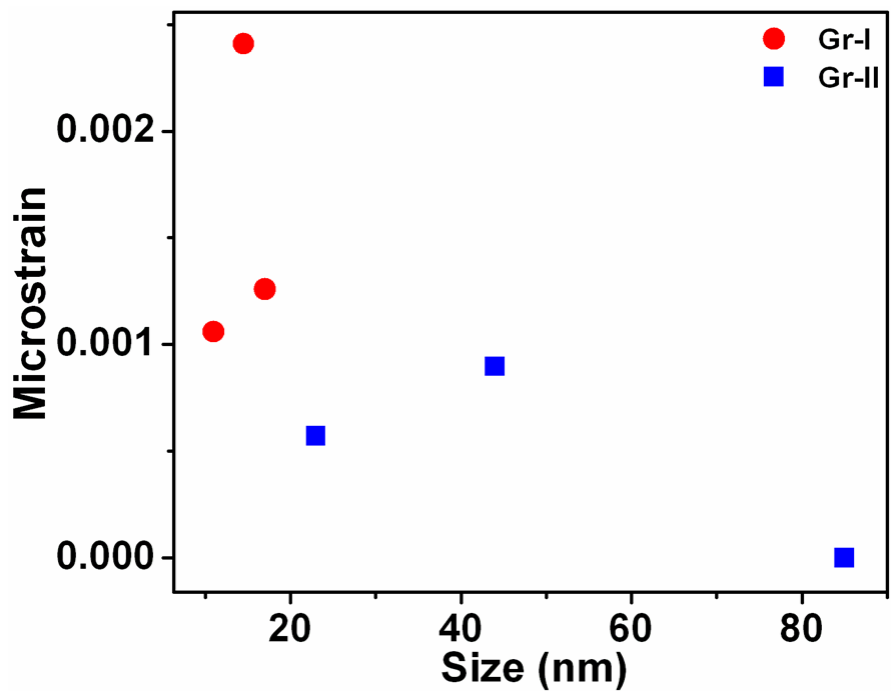

**Figure 3.** Microstrain calculated from XRD data by Williamson-Hall analysis is plotted as function of the size. The relaxation of the strain at the shape transition is apparent.

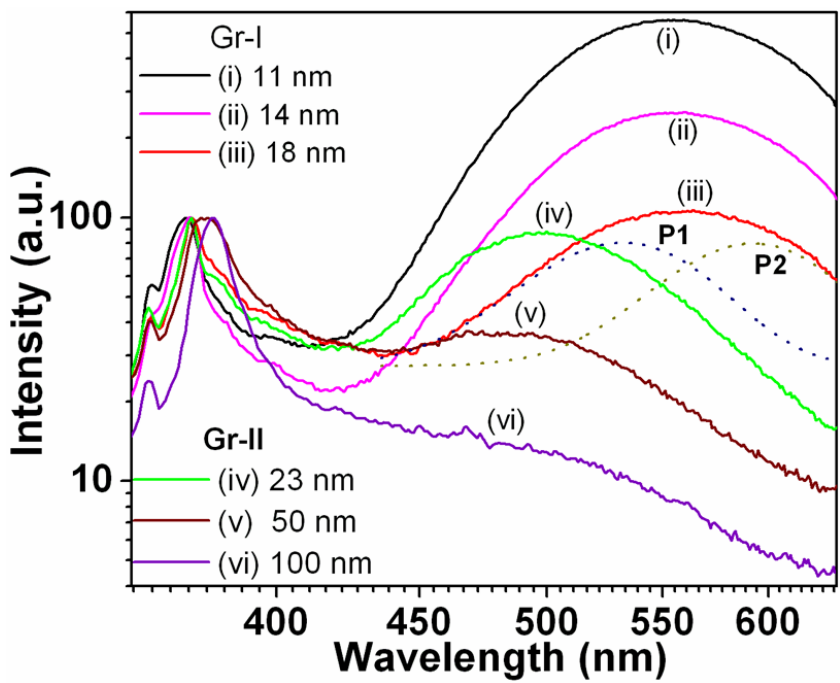

**Figure 4.** The emission bands of the ZnO nanostructures of different size and shape. The excitation was at 325 nm. The sharp dependence of the emission band at 490-565 nm on the size can be seen. A typical emission band is composed of two components P1 and P2 (see text) as shown for the emission curve (iii). The shift in the spectral components when the size increases beyond ~20 nm is prominent.

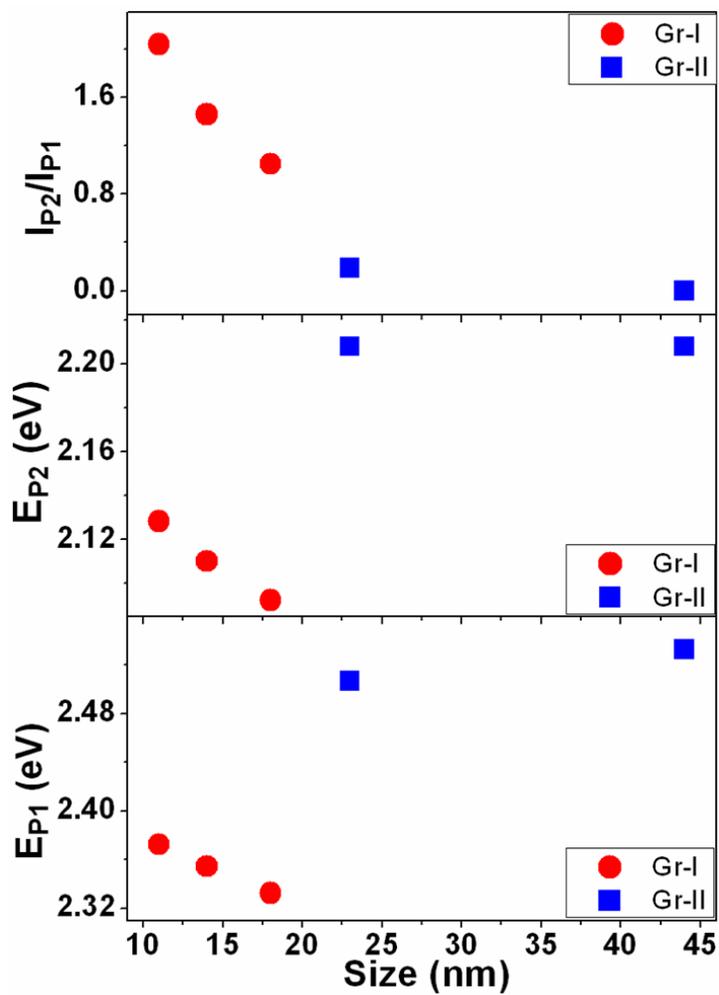

**Figure 5.** Important physical parameters as a function of size: Lower panel: energy of P1 band. Middle panel: energy of P2 band. Upper panel: intensity ratio of P2 to P1 emission bands.

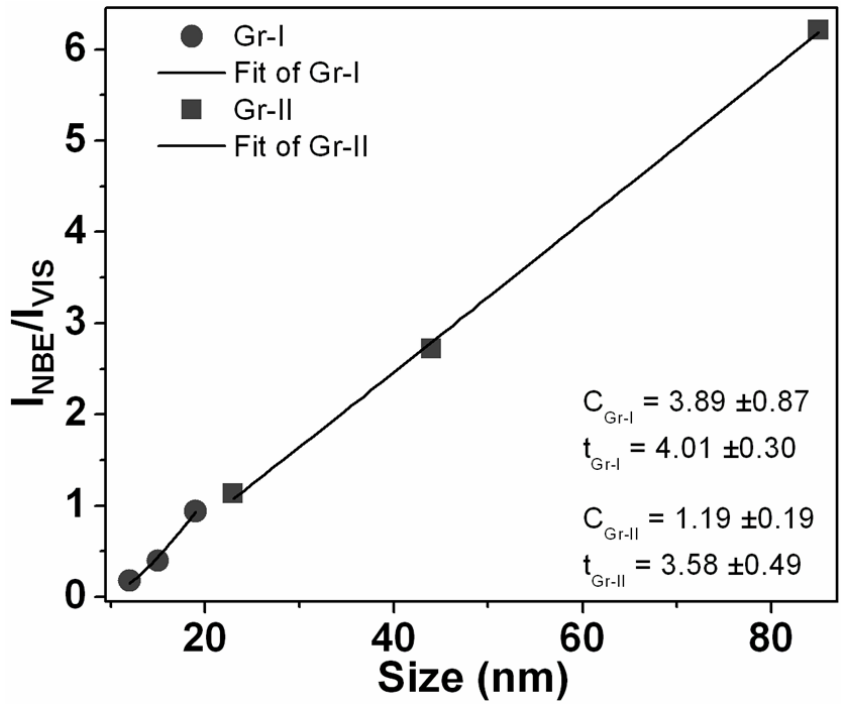

**Figure 6.** Intensity ratio of NBE to visible emission for Gr-I and Gr-II samples. The data are fitted by a surface recombination model with surface layer thickness of 4 nm and 3.6 nm for Gr-I and Gr-II samples respectively.

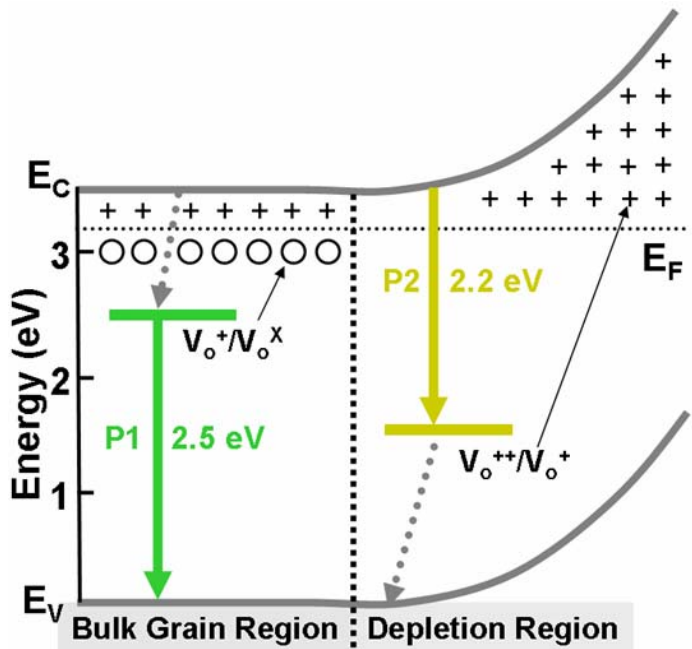

**Figure 7.** Schematic diagram depicting the origin of P1 and P2 emission bands. The defect distributions in the bulk and depletion regions are shown. $E_c$, $E_v$ and $E_F$ are the conduction band, valance band and Fermi energy levels respectively.

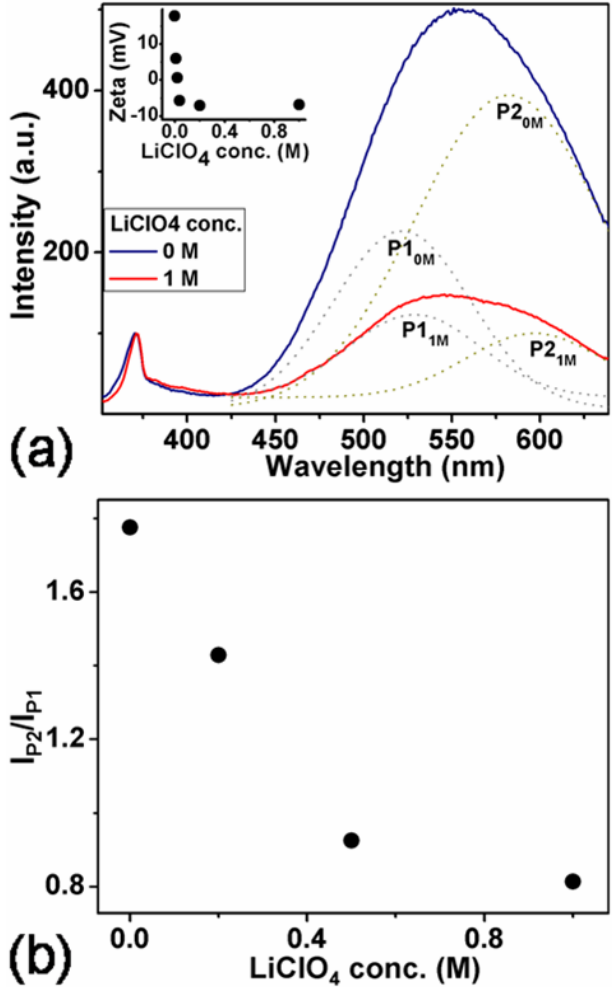

**Figure 8.** (a) The photoluminescence spectra of spherical ZnO nanoparticles (size ~ 11 mn) dispersed in LiClO$_4$ solution in ethanol as indicated on the graph. The two spectral components P1 and P2 have been extracted and displayed for two representative LiClO$_4$ concentrations viz., 0 M and 1 M. Inset shows the zeta potential values of the nanospheres for different concentration of LiClO$_4$ in ethanol. (b) The intensity ratio of the P2 to P1 emission band gradually decreases as the LiClO$_4$ concentration increases.